\documentclass[runningheads]{llncs}

\input{psfig.sty}
\usepackage{graphicx}
\begin{document}

\pagestyle{headings}

\mainmatter

\title{File-based storage of Digital Objects and constituent datastreams: XMLtapes and Internet Archive ARC files}

\titlerunning{Lecture Notes in Computer Science}
\author {Xiaoming Liu\inst{1}
\and Lyudmila Balakireva\inst{1}
\and Patrick Hochstenbach \inst{2} 
\and Herbert Van de Sompel\inst{1}}

\institute{Research Library\\
Los Alamos National Laboratory\\
Los Alamos, NM, US 87544\\
\email{\{liu\_x,ludab, herbertv\}@lanl.gov}\\
\and
University Library\\
Ghent University\\
Rozier 9, B-9000 Ghent Belgium\\
\email{Patrick.Hochstenbach@ugent.be}
}

\maketitle

\begin{abstract}

This paper introduces the write-once/read-many XMLtape/ARC storage approach for Digital Objects and their constituent datastreams. The approach combines two interconnected file-based storage mechanisms that are made accessible in a protocol-based manner. First, XML-based representations of multiple Digital Objects are concatenated into a single file named an XMLtape.  An XMLtape is a valid XML file; its format definition is independent of the choice of the XML-based complex object format by which Digital Objects are represented. The creation of indexes for both the identifier and the creation datetime of the XML-based representation of the Digital Objects facilitates OAI-PMH-based access to Digital Objects stored in an XMLtape. Second, ARC files, as introduced by the Internet Archive, are used to contain the constituent datastreams of the Digital Objects in a concatenated manner. An index for the identifier of the datastream facilitates OpenURL-based access to an ARC file. The interconnection between XMLtapes and ARC files is provided by conveying the identifiers of ARC files associated with an XMLtape as administrative information in the XMLtape, and by including OpenURL references to constituent datastreams of a Digital Object in the XML-based representation of that Digital Object.

\end{abstract}

\section{Introduction and Motivation}

Digital Library architectures that are concerned with long-term access to digital materials face interesting challenges regarding the representation and actual storage of Digital Objects and their constituent datastreams.  With regard to representation of Digital Objects, a trend can be observed that converges on the use of XML-based complex object formats such as the MPEG-21 Digital Item Declaration Language \cite{did2003} or METS \cite{mets}.   In these approaches, the Open Archival Information System \cite{oais} Archival Information Package (OAIS AIP) that represents a Digital Object is an XML-wrapper document that contains a variety of metadata pertaining to the Digital Object, and that provides the constituent datastreams of the Digital Object either By-Value (base64-encoded datastream inline in the XML-wrapper) or By-Reference (pointer to the datastream inline in the XML-wrapper).  This choice for XML is not surprising.  Indeed, both its platform-independence nature and the broad industry support provide some guarantees regarding longevity or, eventually, migration paths.  Moreover, a broad choice of XML processing tools is available, including tools that facilitate the validation of XML documents against schema definitions that specify compliance with regard to both structure and datatypes.

However, the choice of an XML-based AIP format is only part of the solution.  The Digital Objects - represented by means of XML-wrapper documents - and their constituent datastreams still need to be stored.  With this respect, less convergence is observed in Digital Library architectures, and the following approaches have been explored or are actively used:
\begin{itemize}
\item Storage of the XML-wrapper documents as individual files in a file system: On most operating systems, this approach is penalized by poor performance regarding access, and especially back-up/restore.  Also, the OAIS reference model recommends against the storage of Preservation Description Information and Content Information using directory or file-based naming conventions. 
\item
Storage of the XML-wrapper documents in SQL or native XML databases: This approach provides a flexible storage approach, but it raises concerns for long-term storage because, in database systems, the data are crucially dependent on the underlying system. 
\item Storage of the XML-wrapper documents by concatenating many such documents into a single file such as tar, zip, etc.: This approach is appealing because it builds on the simplest possible storage mechanism - a file - and it alleviates the problems of the ``individual file'' approach mentioned before.  However, off-the-shelf XML tools are not efficient to retrieve individual XML-wrapper documents from such a concatenation file. 
\end{itemize}

The Internet Archive has devised the ARC file \cite{arc}, a file-based approach to store the datastreams that result from Web crawling.  In essence, an ARC file is the concatenation of many datastreams, whereby a separation between datastreams is created by a section that provides -- mainly crawling-related -- metadata in a text-only format.  Indexing tools are available to allow rapid access to datastreams based on their identifiers.  While the file-based approach to store a collection of datastreams is attractive, the ARC file format has limited capabilities for expressing metadata.  Even the result of the ongoing revision of the ARC file format, in which the authors are involved, will probably not allow expressing the extensive metadata that is typical for Archival Information Packages in Digital Libraries.  By all means, it is not clear how various constituent datastreams of a Digital Object could be tied together in the ARC file format, or how their structural relationships could be expressed.  Moreover, ARC files do not provide the validation capabilities that are part of what makes XML-based representation and storage attractive.

In this paper we introduce a representation and storage approach for Digital Objects that was pioneered in the aDORe repository effort of the Research Library of the Los Alamos National Laboratory (LANL).  The approach combines the attractive features of the aforementioned techniques by building on two interconnected file-based storage approaches, XMLtapes and ARC files. These file formats are proposed as a long-term storage mechanism for Digital Objects and their constituent datastreams.  The proposed storage mechanism is independent of the choice of an XML-based complex object format to represent Digital Objects. It is also independent of the indexing technologies that are used to access embedded Digital Objects or constituent datastreams: as technologies evolve, new indexing mechanisms can be introduced, while the file-based storage mechanism itself remains unchanged.

\section{Representing Digital Objects}
Over the last 2 years, the Digital Library Research and Prototyping Team of the LANL Research Library has worked on the design of the aDORe repository architecture \cite{adore} aimed at ingesting, storing, and making accessible to downstream applications a multi-TB heterogeneous collection of digital scholarly assets. 

As is the case in most Digital Libraries, assets stored in aDORe are {\it complex} in the sense that they consist of multiple individual datastreams that jointly form a single logical unit.  That logical unit can, for example, be a scholarly publication that consists of a research paper in PDF format, metadata describing the paper expressed in XML, and auxiliary datastreams such as images and videos in various formats, including TIFF, JPEG and MPEG.  For reasons of clarity, this paper will refer to an asset as a {\it Digital Object}, and to the individual datastreams of which the asset consists as {\it constituent datastreams}.  The complex nature of the assets to be ingested into aDORe led to an interest in representing assets by means of XML wrappers, which itself resulted in the selection of the MPEG-21 DIDL as the sole way to represent the asset by means of XML documents called DIDL documents.  The actual use of the MPEG-21 DIDL in aDORe is described in some detail in the slightly outdated \cite{didl:dlib} and the more recent \cite{didl:jdl}.  Although this paper will illustrate the XMLtape/ARC storage mechanism for the case where MPEG-21 DIDL is used to represent Digital Objects, it will become clear that the approach is independent of the choice of a specific XML-based complex object format.  Hence, it could also be used when Digital Objects are represented using METS or IMS/CP \cite{ims}.  

An important, OAIS-inspired, characteristic of the aDORe environment is its write-once/read-many strategy.  Indeed, whenever a new version of a previously ingested Digital Object needs to be ingested, a new DIDL document is created; existing DIDL documents are never updated or edited.  The distinction between multiple versions of a Digital Object is achieved through the use of 2 types of identifiers that are clearly recognizable, and expressed as URIs in DIDL documents:

\begin{description}
\item[Content Identifiers] Content Identifiers corresponds to what the OAIS categorizes as Content Information Identifiers. Content Identifiers are directly related to identifiers that are natively attached to Digital Objects before their ingestion into aDORe.  Indeed, in many cases such Digital Objects, or their constituent datastreams, have identifiers that were associated with them when they were created or published, such as Digital Object Identifiers \cite{doi} for scholarly papers.  Different versions of a Digital Object have the same Content Identifier.
\item[Package Identifiers]  A DIDL document that represents a Digital Object functions as an OAIS AIP in aDORe.  During the ingestion process, this DIDL document itself is accorded a globally unique identifier, which the OAIS categorizes as an AIP Identifier. Values for Package Identifier are constructed using the UUID algorithm \cite{uuid}; they are expressed as URIs in a reserved sub-namespaces of the 'info:lanl-repo/' namespace, which the LANL Research Library has registered under the info URI Scheme \cite{info}.   
\end{description}

A separate component in the aDORe architecture, the Identifier Locator, keeps track of all versions of a Digital Object.  

\section{Storing and accessing Digital Objects in XMLtapes and ARC files}
The aDORe environment shares two important characteristics with the Internet Archive:
\begin{itemize}
\item Data feeds in aDORe are typically received in batches, each of which can contain anywhere between 1,000 and 1,000,000 Digital Objects.
\item Ingestion of a previously ingested Digital Object does not result in editing of that previously ingested version, but rather to a from-scratch ingestion of the new version.
\end{itemize}
These characteristics suggest that a file-based, write-once/read-many storage approach should be as appealing to aDORe as it is to the Internet Archive.  However, Internet Archive ARC files have only limited capabilities to express metadata pertaining to datastreams and to the ingestion process, and they have no obvious way to express structure of a Digital Object with multiple constituent datastreams.  Therefore, in aDORe, an approach has been devised that combines two interconnected file-based storage mechanisms: XMLtapes and ARC files.

\subsection{XMLtapes: File storage of XML-based representations of Digital Objects}
An XMLtape is an XML file that concatenates the XML-based representation of multiple Digital Objects.  In the aDORe implementation of the XMLtape, the XML-based representations of Digital Objects are DIDL documents compliant with the MPEG-21 DIDL standard.  In order to keep these DIDL documents small and hence easy to process, they typically contain:
\begin{description}
\item[By-Value] The metadata pertaining to the Digital Object, its constituent datastreams, and the ingestion process. 
\item[By-Reference] The constituent datastreams of the represented Digital Object.  The embedded reference in the DIDL document points to the datastream that is stored in an ARC file that is associated with the XMLtape.  The nature of the reference and the access mechanism will be explained in Section 3.3 and Section 3.4, respectively.
\end{description}
The structure of XMLtapes is defined by means of an XML Schema \cite{tape:schema}:
\begin{itemize}
\item An XMLtape starts off with a section that allows for the inclusion of administrative information pertaining to the XMLtape itself.  Typical information includes provenance information of the contained batch of Digital Objects, identification of the processing software, processing time, etc.  
\item The XMLtape-level administrative section is followed by the concatenation of records, each of which has administrative information attached to it.  While allowing for the inclusion of a variety of record-level administrative information, the XMLtape has two strictly defined administrative elements: the identifier and creation datetime of the contained record.  This allows for the use of a generic XMLtape processing tool that is independent of the nature of the actual included records.  In aDORe, these strictly defined administrative information elements translate to the Package Identifier and the creation datetime of the DIDL document that is a record in the XMLtape. 
\item The records provided in an XMLtape can be from any XML Namespace.  In aDORe, they are DIDL documents compliant with the MPEG-21 DIDL XML Schema \cite{didl:schema}.
\item The XMLtape itself is a valid and well-formed XML file that can be handled by off-the-shelf XML tools for validation or parsing.
\end{itemize}

In order to interpret an XML file, it is generally necessary to parse and load the complete file.  In case of XMLtapes, such an approach would forbid fast retrieval of the embedded XML documents.  Therefore, in order to optimize access, two indexes are created.   The indexes correspond with the mandatory record-level administrative information, and have identifier and creation datetime of the embedded records as their respective keys.  As will be explained in Section 3.5, these indexes facilitate OAI-PMH access to the XML documents contained in the XMLtape.  In addition to these identifier and datetime keys, each index stores the byte-offset and byte-count per matching record.   When retrieving a record from an XMLtape, first a lookup in an index file is required to fetch a record position, followed by a seek into the XMLtape to return the required record.  

\subsection{ARC files: File storage of constituent datastreams of Digital Objects}

In some scenarios, it can make sense to physically embed certain constituent datastreams of a Digital Object in the DIDL document that is contained in an XMLtape. For example, embedding descriptive metadata or image thumbnails may improve access speed for downstream applications. However, in other scenarios, such embedding is neither optimal nor realistic. Indeed, the mere size of a constituent datastream, worsened by the required base64 encoding, leads to large DIDL documents that may cause serious XML processing challenges at the time of dissemination.  

The ARC file format \cite{arc} is used by Internet Archive to store datastreams resulting from large-scale web crawling.  The ARC file format is structured as follows:
\begin{itemize}
\item An ARC file has a file header that provides administrative information about the ARC file itself. 
\item The file header is followed by a sequence of document records.  Each such record starts with a header line containing some, mainly crawl-related, metadata.  The most important fields of the header line are the URI of the crawled document, the timestamp of acquisition of the data, and the size of the data block that follows the header line. The header line is followed by the response to a protocol request such as an HTTP GET.  
\end{itemize}

Tools such as those from netarchive.dk \cite{netarchive} are available to generate and consult an index external to the ARC file that facilitates rapid access to contained records, using their URI as the key.  As will become clear from Section 3.3, these tools play a core role when connecting XMLtapes and associated ARC files.

\subsection{Associating ARC files with an XMLtape during the ingestion process}

Both the XMLtape and its associated ARC files are created during the ingestion process.  An insight in the ingestion flow is given here:
\begin{itemize}
\item When a feed of Digital Objects is obtained from an information provider, the ingestion process creates a DIDL document per obtained Digital Object; each DIDL document receives a globally unique Package Identifier.  
\item Typically, all DIDL documents for a given batch are stored in a single XMLtape that can easily store over 1,000,000 DIDL documents.  An XMLtape itself also receives a globally unique XMLtape Identifier.
\item Depending on the size of the constituent datastreams of the Digital Objects in a given feed, one or more ARC files are created during the ingestion process.  Each ARC file is given a globally unique ARC file Identifier, and ARC files are associated with the XMLtape by including these ARC file Identifiers in the XMLtape-level administrative section.
\item For each DIDL document written to the XMLtape:
\begin{itemize}
\item Each constituent datastream of the represented Digital Object is accorded a globally unique Datastream Identifier that has no relation to the aforementioned Package Identifiers or Content Identifiers.  
\item The constituent datastream is written to an ARC file; the URI field of the ARC file record header receives the Datastream Identifier as its value.  
\item A reference to the constituent datastream is written in the DIDL document.  Core elements in this reference are the ARC file Identifier and the Datastream Identifier.  As will be explained in the next Section, these references are encoded in a manner compliant with the NISO OpenURL standard \cite{openurl}.
\end{itemize}
\item Indexes are created for both the XMLtape and its associated ARC files:
\begin{itemize}
\item For the XMLtape, two indexes are created, with the Package Identifier and the creation datetime of DIDL documents as their respective keys.
\item Per ARC file, an index is created that has the Datastream Identifier as its key.
\end{itemize}
\item All globally unique identifiers accorded during the ingestion process are created based on the UUID algorithm \cite{uuid}.
\end{itemize}

\subsection{Adding protocol-based access to XMLtapes and ARC files}

The features described in the previous Sections allow for a persistent standards-based access to both file-based storage mechanisms.

Each XMLtape is exposed as an autonomous OAI-PMH \cite{oai} repository with the following characteristics (Protocol elements are shown in \textbf{bold}): 
\begin{itemize}
\item It has a  \textbf{baseURL(XMLtape)}, which is an HTTP address that contains the XMLtape Identifier to ensure its uniqueness.
\item Contained \textbf{records} are DIDL documents only.
\item The \textbf{identifier} used by the OAI-PMH is the Package Identifier.
\item The \textbf{datestamp} used by the OAI-PMH is the creation datetime of the DIDL document.
\item Access based on \textbf{identifier} and \textbf{datestamp} is enabled via the 2 aforementioned indexes created per XMLtape.
\item The only supported metadata format is DIDL, with \textbf{metadataPrefix} DIDL, defined by the MPEG-21 DIDL XML Schema. 
\item The supported OAI-PMH harvesting \textbf{granularity} is seconds-level.
\end{itemize}

Each ARC file is exposed as an OpenURL Resolver:  
\begin{itemize}
\item The OpenURL Resolver has a \textbf{baseURL(ARC file)}, which is an HTTP address that contains the ARC file Identifier to ensure its uniqueness. 
\item References embedded in the DIDL documents are compliant with the NISO OpenURL standard \cite{openurl}.  As a matter of fact, the reference uses the HTTP-based, Key/Encoded-Value Inline OpenURL.  The Referent of this OpenURL is a datastream stored in an ARC file, and this datastream is described on the OpenURL by means of its Datastream Identifier.  
\item The sole service provided by the OpenURL Resolver is delivery of the datastream referenced on the OpenURL.  This service is enabled by the index that is created per ARC file.
\end{itemize}
\subsection{Accessing Digital Objects and constituent datastreams}
\begin{figure}[h]
\centerline{
\includegraphics[width=5in]{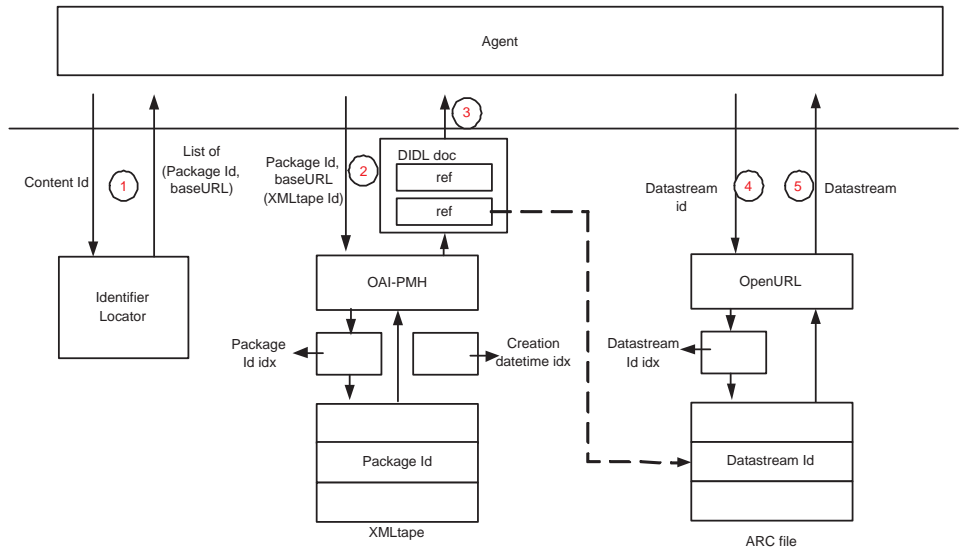}
}
\caption{Accessing a Digital Object stored in  XMLtape/ARC}
\label{fig:fig1}
\end{figure}

This Section explains how Digital Objects and constituent datastreams are accessed in the aDORe environment in which the XMLtape/ARC approach is used as file-based storage mechanism.  Figure \ref{fig:fig1} is provided to support a better understanding of the flow.  In what follows, protocol elements are shown in \textbf{bold}, while argument values are shown in {\it italic}.
\begin{itemize}
\item In a typical scenario, an agent requests a Digital Object from aDORe by means of its Content Identifier.  The Identifier Locator, not described in this paper, contains information on the locations of all version of a Digital Object with a given Content Identifier.  When queried, it returns a list of Package Identifiers of DIDL documents that represent the given Digital Object, and for each returned Package Identifier the OAI-PMH \textbf{baseURL}(XMLtape Identifier) of the XMLtape in which the DIDL document resides.
\item Next, the requesting agent selects a specific version of a Digital Object, thereby implicitly selecting a specific XMLtape Identifier and the OAI-PMH \textbf{baseURL}(XMLtape Identifier) of the XMLtape in which the chosen DIDL document resides.  This DIDL document can be obtained using the OAI-PMH request:
\\
\begin{tabular}{ l l }
\textbf{baseURL}(XMLtape Identifier)?&\textbf{verb=GetRecord} \&\\
&\textbf{identifier=}{\it Package Identifier}\&\\
&\textbf{metadataPrefix=}{\it didl}\\
\end{tabular}

\item Issuing this OAI-PMH request results in a look-up of the Package Identifier in the identifier-based index that was created for the targeted XMLtape.  This look-up reveals the byte-offset and byte-count of the required DIDL document in the XMLtape.  Given this information, a process can access the DIDL document in the XMLtape and return it to the agent.
\item Having obtained a representation of the requested Digital Object, the requesting agent may decide to next request a constituent datastream.  Because such datastreams are included in the DIDL document By-Reference, selection of a specific datastream is equivalent to selecting the OpenURL that references it.  This OpenURL is as follows:
\\
\begin{tabular}{ l l }
\textbf{baseURL}(ARC file Identifier)?&\textbf{url\_ver=Z39.88-2004 }\&\\
&\textbf{rft\_id=}{\it Datastream Identifier}\\
\end{tabular}

\item Issuing this OpenURL request results in a look-up of the Datastream Identifier in the URI-based index that was created for the targeted ARC file.  This look-up reveals the byte-offset and byte-count of the required datastream in the ARC file.  Given this information, a process can access the datastream in the ARC file and return it to the agent.
\end{itemize}

\subsection{Implementation}

All XMLtape and ARC file components are implemented in Java.   Due to the standards-based approach, several off-the-shelf components have been used.  The XMLtape indexes are implemented with Berkeley DB Java Edition \cite{berkeleydb}, while OAI-PMH access is facilitated by OCLC's OAICat software \cite{oaicat} which, in collaboration with OCLC, was extended to support access to multiple OAI-PMH repositories in a single installation.  Creation, indexing and access to ARC files are implemented using the netarchiv.dk toolset \cite{netarchive}.

The performance and scalability of ARC files are demonstrated by the Internet Archive and its Wayback Machine, which stores more than 400 terabytes of data. The performance of the XMLtape solution depends on the choice of the underlying indexing and retrieval tools. The file-based nature of both XMLtapes and ARC files makes it straightforward to distribute content over multiple disks and servers.
\section{Future Work}
Two aspects of the reported work will require future updating of the XMLtape/ARC approach:
\begin{itemize}
\item 
First, a problem related to the indexing of XMLtapes must be resolved. Many XML parsers do not support byte-level processing. However, correct byte-level location is essential to yield a waterproof solution for the two indexes that are created for XMLtapes, both of which are based on byte-count and byte-offset. This problem currently limits the choice of XML tools that can be used for the indexing process.  A fundamental solution to this problem should come from support for the DOM Level 3 API \cite{dom3} in XML tools, as this API requires support for byte-level location.

\item Second, under the umbrella of the International Internet Preservation Consortium (IIPC) \cite{iipc}, a conglomerate of the Internet Archive and national libraries, the ARC file format is undergoing a revision.  Formal requirements for the revised format have been specified, including OAIS compliance, ability to deal with all Internet protocols, support of metadata, and capability to verify data integrity \cite{arcrevision}.  The authors are involved in this effort, and have provided input, some of which is aimed at making the revised file format even more suitable for the use case of storing local content, in addition to the typical Web crawling use case \cite{arccomment}.  At the time of writing, a draft proposal for a WArc file format is available and awaiting further comments.  Once a new format is accepted, existing ARC files in aDORe will be converted, and new tools compliant with the new format will be put in place.
\end{itemize}
\section{Conclusions}

This paper has described a storage approach for Digital Objects and its constituent datastreams that has been pioneered in the context of the aDORe repository effort by the LANL Research Library.  The approach combines two interconnected file-based storage mechanisms that are made accessible in a protocol-based manner:
\begin{itemize}
\item XMLtapes concatenate XML-based representations of multiple Digital Objects, and are made accessible through the OAI-PMH.  
\item ARC files concatenate constituent datastreams of Digital Objects, and are made accessible through an OpenURL Resolver.  
\item The interconnection between both is provided by conveying the identifiers of ARC files associated with an XMLtape as administrative information in the XMLtape, and by including OpenURL references to constituent datastreams of a Digital Object in the XML-based representation of that Digital Object.
\end{itemize}

The approach is appealing for several reasons:
\begin{itemize}
\item The file-based approach is inherently simple, and dramatically reduces the dependency on other components as it exists with database-oriented storage.  
\item The disconnection of the indexes required for access from the file-based storage approach allows retaining the files over time, while the indexes can be created using other techniques as technologies evolve.  
\item The protocol-based nature of the access further increases the flexibility in light of evolving technologies as it introduces a layer of abstraction between the access method and the technology by which actual access is implemented.
\item The XMLtape approach is inspired by the ARC file format, but provides several additional attractive features.  It provides a native mechanism to store XML-based representations of Digital Objects that are increasingly being used in Digital Library architectures.  This yields the ability to use of off-the-shelf XML processing tools for tasks such as validating and parsing.  It also provides the flexibility to easily deal with Digital Objects that have multiple constituent datastreams, and to attach a wide variety of metadata to both those Digital Objects and their datastreams.  Of special interest for preservation purposes is the ability to include XML Signatures for constituent datastreams (stored themselves outside of the XMLtape) as metadata within the XML-based representation of a Digital Object stored in the XMLtape.
\item Used in this dual file-based storage approach, ARC files keep the appeal they have in the context of the Internet Archive.  For aDORe, they are appealing for additional reasons, including the existence of off-the-shelf processing tools, the proven use in a large-scale environment, and the prospect of the format -- or a new version thereof -- being used in the international context of the International Internet Preservation Consortium that groups the Internet Archive and national libraries worldwide.
\end{itemize}

As can be understood, the proposed XMLtape/ARC approach is not tied to aDORe's choice of MPEG-21 DIDL as the complex object format to represent Digital Objects.  The approach can also be used when Digital Objects are represented using other formats such as METS or IMS/CP.  As a matter of fact, at LANL, the XMLtape approach is even used to store the results from OAI-PMH harvesting of Dublin Core records, in which case the record-level administrative information contains the OAI-PMH identifier and datestamp of the Dublin Core record to which it is attached.  While currently untested, the proposed approach could also be used as a mechanism to transport large archives encoded as XMLtape/ARC collections from one system to another.  

\section{Acknowledgments}

The authors would like to thank Jeff Young from OCLC for his willingness to update the OAICat software to accommodate the multiple repository use case.  And many thanks to our LANL Research Library colleagues Jeroen Bekaert, Mariella Di Giacomo, and Thorsten Schwander for their input in devising many facets of the aDORe architecture.  Finally thanks to Michael L. Nelson for proofreading a draft of this paper. 

The reported work is partially funded by an NDIIP grant from the Library of Congress.

\end{document}